\documentclass[epj]{svjour}
\usepackage{graphics,color}
\newcommand{\CBB}{}

\newcommand{\Cmm}{}
\newcommand{\Cm}{}
\newcommand{\Black}{}
\newcommand{\gray}{}
\newcommand{\urple}{}
\newcommand{\Red}{}

\newcommand{\Violet}{}

\newcommand{\treintaycinco}{{\bf 35}}
\newcommand{\cincuentayseis}{{\bf 56}}
\newcommand{\setenta}{{\bf 70}}

\newcommand{\setecientos}{{\bf 700}}
\newcommand{\milcientotreintaycuatro}{{\bf 1134}}

\newcommand{\lambdd}{{\bar\lambda}}
\begin{document}
\title{Meson-Baryon s--wave Resonances with Strangeness --3}
\author{C. Garc{\'\i}a-Recio \and J. Nieves \and
L. L. Salcedo }
%
\offprints{}          
\institute{Departamento de 
F{\'\i}sica At\'omica, Molecular y Nuclear,
Universidad de Granada, E-18071 Granada, Spain}
\date{Received: date / Revised version: date}
%
\abstract{Starting from a consistent SU(6) extension of the Weinberg-Tomozawa
  (WT) meson-baryon chiral Lagrangian (Phys. Rev.  \textbf{D74} (2006)
  034025), we study the s--wave meson-baryon resonances in the strangeness
  $S=-3$ and negative parity sectors. Those resonances are generated by
  solving the Bethe--Salpeter equation with the WT interaction used as kernel.
  The considered mesons are those of the {\bf 35}-SU(6)-plet, which includes
  the pseudoscalar (PS) octet of pions and the vector (V) nonet of the rho
  meson.  For baryons we consider the {\bf 56}-SU(6)-plet, made of the $1/2^+$
  octet of the nucleon and the $3/2^+$ decuplet of the Delta.  Quantum numbers
  $I(J^P)=0(3/2^-)$ are suggested for the experimental resonances
  $\Omega^*(2250)^-$ and $\Omega^*(2380)^-$.  Among other, resonances with
  $I=1$ are found, which minimal quark content is $sss\bar{l}l'$, being $s$
  the $strange$ quark and $l,~l'$ any of the the light $up$ or $down$ quarks.
  A clear signal for such a pentaquark would be a baryonic resonance with
  strangeness $-3$ and electric charge of $-2$ or $0$, in proton charge units.
  We suggest looking for $K^-\Xi^-$ resonances with masses around 2100 and
  2240~MeV in the sector $1(1/2^-)$, and for $\pi^{\pm}\Omega^-$ and
  $K^-{\Xi^*}^-$ resonances with masses around 2260~MeV in the sector
  $1(3/2^-)$.  }
%
\maketitle
\section{Introduction}
\label{sect:Introduction}

Using a spin-flavor-SU(6) extended Weinberg\--Tomozawa (WT) meson-baryon
interaction\footnote{This WT interaction has also been extended to arbitrary
  number of colors and flavors in ref.~\cite{CGR:Nc}}~\cite{CGR:WT6}, we study
the s-wave resonances with strangeness $S=-3$, isospin I=$0,1$ and spin-parity 
J$^P$ = 1/2$^-$, 3/2$^-$, 5/2$^-$.
The resonances are generated by solving the Bethe--Salpeter equation with the
extended WT meson-baryon interaction used as a kernel.
In this model, the involved mesons are those of the \Cmm{{\bf
    35}-SU(6)-plet}=\Cm{$8_1$}$\oplus$\Cm{$8_3\oplus 1_3$}, which includes
the PS meson octet of the pions, \Cm{($\pi$, $\eta$, $K$, $\bar{K})\in 8_1$}
and the V nonet of the rho meson, \Cm{($\rho$, $\omega$, $\phi$, $K^*$,
  ${\bar K^*}$)$\in 8_3\oplus 1_3$}.
We approximate the $\eta$ meson as the isospin singlet state of the PS
SU(3)-octet. For the $\omega$ and $\phi$ vector mesons, we assume ideal mixing
among the V singlet and octet mesons.
The baryons are those of the {\bf 56}-SU(6)-plet=$8_2\oplus 10_4$, which
contains the 1/2$^+$ octet of the N ($N,~\Lambda,~\Sigma,~\Xi$) and the
3/2$^+$ decuplet of the $\Delta$ ($\Delta,~\Sigma^*,~\Xi^*,~\Omega$).
Masses, widths and couplings of the resonances found are calculated and, when
possible, comparison with experimental ones is attempted.

Unitary extensions of chiral perturbation theory to study meson-baryon
interactions using a coupled channel scheme were introduced some time
ago~\cite{previousChBS}. They have been successfully applied in the theoretical
microscopical description of meson-baryon scattering and of well known lowest
lying baryon resonances, which were shown to be dynamically generated. Thus
different $J^P=1/2^-$ $s-$wave resonances ( $\subset$~8$_{\pi}\times$8$_N$,
made of PS mesons of the pion octet and of the baryons of the nucleon octet)
like $N^*(1535)$, $\Lambda(1405)$, $\Lambda(1670)$, $\Sigma(1620)$ and 
$\Xi(1620)$~\cite{sucessfulSU3a} and, more recently, the $J^P=3/2^-$
$d-$wave resonances ( $\subset$~8$_{\pi}\times$10$_\Delta$, made of PS mesons
of the pion octet and of the baryons of the delta decuplet) like
$\Lambda(1520)$, $\Sigma(1670)$ and $\Xi(1820)$~\cite{sucessfulSU3b} have been
found and their properties studied. Those previous chiral Bethe-Salpeter
coupled channels unitary approaches using the WT kernel have included hadron
multiplets belonging to the flavor SU(3) irreducible representations. In the
case of mesons, the only ingredient has been the octet of PS mesons.
\section{Model}
Motivations for extending the previous SU(3) based models to a SU(6) extended
model are the following. First, in the large $N_c$ limit, $8_N$ and
$10_\Delta$ are degenerated and form a 56-multiplet of spin--flavor SU(6).
Second, vector mesons do exist, interact and couple to baryons. Third, there
are baryonic resonances decaying to a PS meson and a baryon, but also to a V
meson and a baryon, for instance, the strangeness --3 resonance
${\Omega^*(2380)}^-$ decays to $K^-{\Xi^*}^0$ and to ${\bar K}^*{}^0 \Xi^-$
with similar strengths and of the same order as the other known decay mode
($\bar K^-\pi^0\Xi^-$)~\cite{Biagi:1986}. All of these call for a 'SU(6)'
model which deals all together with the 56-baryons and 35-mesons like that of
ref.\cite{CGR:WT6}. We consider this spin--flavor symmetric scenario as a
reasonable first approach. In the 'SU(6)' approach the interacting kernel
$V^{IYJ}$ for a sector with hypercharge $Y$ (strangeness plus one for
baryons), isospin $I$ and spin $J$ is given by:
\begin{eqnarray}
\langle \Cmm{m_i},~\CBB{B_i} |V^{IYJ}(s)|\Cmm{m_j},~\CBB{B_j}\rangle& 
= &\Violet{D}^{IYJ}_{ij}~
\frac{2\,\sqrt{s}-M_{\CBB{B_i}}-M_{\CBB{B_j}}}{4\,f^2} 
\nonumber
\\
\Cmm{m_i},\Cmm{m_j}\in\Cmm{\treintaycinco} ,~
\CBB{B_i},\CBB{B_j}\in\CBB{\cincuentayseis}
,& &
\Violet{D}^{IYJ} = \sum_{\nu}
{\lambdd}_{\nu} ~ \hat P^{IYJ}_{\nu}
\nonumber
\\
\lambdd_{\cincuentayseis}= -12,~
\lambdd_{\setenta}=-18,& & 
\lambdd_{\setecientos}=6,~
\lambdd_{\milcientotreintaycuatro} =-2,
\nonumber
\end{eqnarray}
%
$\hat P^{IYJ}_{\nu}$ is the projector of the meson-baryon into the
$\nu$--SU(6)-representation, and $\nu$ runs over 
$\cincuentayseis$, $\setenta$, $\setecientos$, $\milcientotreintaycuatro$ 
 because 
 $\treintaycinco$$\otimes$$\cincuentayseis$= $\cincuentayseis
 \oplus\setenta\oplus\setecientos\oplus\milcientotreintaycuatro$.\\
The $\lambdd_\nu$ positive (negative) means that in channel $\nu$ the
meson-baryon interaction is repulsive (attractive).
When this kernel is restricted to $m_i,m_j$ being only PS mesons, it coincides
with the 'SU(3)' lowest order WT kernels previously used in
refs.\cite{sucessfulSU3a,sucessfulSU3b}.\\
Note that $D$ is SU(6) invariant, this symmetry being
explicitly broken by the different masses of baryons and mesons.
In addition we replace 
the $f^2$ of the interaction by $f_{m_i} f_{m_j}$ for
the $ij$ matrix element. We use $f_\pi= 92.4$~MeV, $f_K=1.15 f_\pi$,
$f_\eta=1.2 f_\pi$~\cite{PDG:2006}, and $ f_{K^*}=f_{K}$, $f_\rho=f_\pi$,
$f_\omega=f_\phi=f_\eta$.
This interaction is used to solve the Bethe--Salpeter coupled channel equation for
the meson-baryon $T-$matrix
\begin{equation}
T ^{-1}(s)~=~V^{-1}(s)~ - ~ J(s)
\end{equation}
where $J(s)$ is the diagonal matrix of the meson-baryon loop
functions~\cite{sucessfulSU3a,sucessfulSU3b}.  
For each channel ($m_i$, $B_i$) it is ultraviolet 
regularized by subtracting a constant so that\\
$J\left(s={m_{m_i}}^2+{M_{B_i}}^2\right)=0.$\\
We test that the inclusion of the new 'SU(6)' channels (those involving V
mesons not included in the 'SU(3)' calculations) does not spoil previous
results which were successful. See ref.~\cite{JuanQNP06} for comparison in the
 sector ($S=-1,~I=0,~J^P=1/2^-$).
\section{Results}
\label{sec:Results}
%
We solve the coupled--channel Bethe--Salpeter equation and  look for  the
$T-$matrix  poles in the second Riemann sheet. Close to a pole the
$T-$matrix behaves as
\begin{equation}
T_{ij}\sim {{g_i g_j}\over{\sqrt{s}-(M_R-i\Gamma_R/2)}}
\end{equation}
and the position of the pole and its residue define the mass $M_R$, width
$\Gamma_R$ and complex coupling constants $g_i$ to different $i$ channels of
the found resonance.

All the calculations are done neglecting the widths of the baryons of the
decuplet and of the V mesons.
When the orbital angular momentum of the meson-baryon system is
zero, the odd parity $S=-3$ resonances formed by coupling 
{\bf 35}-mesons to {\bf 56}-baryons can have the following $I(J^P)$ quantum
numbers: $0(1/2^-)$, $0(3/2^-)$, $0(5/2^-)$, $1(1/2^-)$, $1(3/2^-)$ and
$1(5/2^-)$. The mass, width and absolute values of the coupling constants of
the resonances found for each of those sectors are shown in
tables~\ref{tab:1}, \ref{tab:2}, \ref{tab:3}, \ref{tab:4}, \ref{tab:5} and
\ref{tab:6}, respectively. Resonances with width below
125~MeV are displayed in fig.~1.

Several ($S=-3$, $I=1$) resonances have been found (tables~\ref{tab:4},
\ref{tab:5}
and \ref{tab:6} and full symbols in fig.~1).
These resonances have, at least, three $s$ quarks to provide strangeness
$S=-3$ and a pair of light ($u$ or $d$) quark-antiquark to achive isospin $I=1$
quantum number. Hence, those dynamically generated ($S=-3$, $I=1$) resonances
are pentaquarks and in principle they could be empirically spotted very
easily.  A clear signal of them would be a resonance with strangeness $S=-3$
and electric charge $Q=-2$ (this is $I_z=-1$) with minimal quark content 
$sssd\bar u$. Another clear signature would be a $S=-3$ and $Q=0$ resonance
with a minimal quark content of $sssu\bar d$. From the tables, the best ways
for observing pentaquarks would be, in the sector $I(J^P)=1(1/2^-)$ looking
for $K^-\Xi^-$ resonances with masses around 2100 and 2240~MeV, and in the
sector $I(J^P)=1(3/2^-)$ looking for $\pi^{\pm}\Omega^-$ and $K^-{\Xi^*}^-$
resonances with masses around
2260~MeV.\\
All the experimentally known $\Omega^*$ resonances are listed in
table~\ref{tab:exp}. Tentatively, we identify the experimental isoscalar
$\Omega^*(2250)^-$ to the theoretical $\Omega^*(2265)^-$ of sector $0(3/2^-)$,
because of the relative closeness of masses and widths and, also, of observed
decay channels.  Likewise, the experimental $\Omega^*(2380)^-$ could be
assigned to the found $\Omega^*(2343)^-$ of sector $0(3/2^-)$. We do not find
a clear assignment for the experimental $\Omega^*(2470)$ because it decays to
$\Omega^-\pi^+\pi^-$, and the resonances that we find around that energy have
not the possibility of decaying to such channel.

More details will be given elsewhere.
\begin{figure}[h]
\label{fig:1}       
\resizebox{0.5\textwidth}{!}{%
  \includegraphics{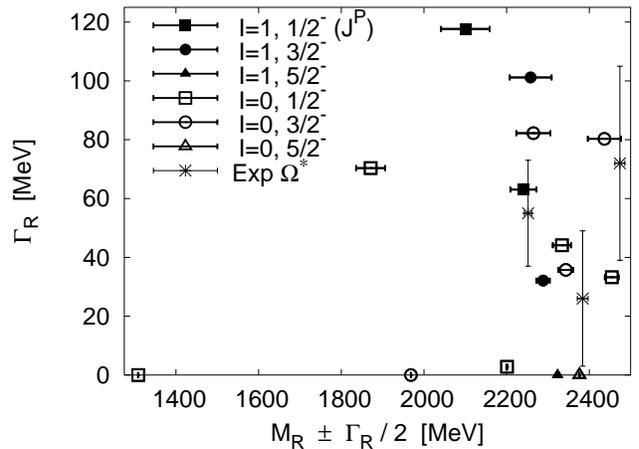}
}
\protect{\caption{Spin--parity $J^P=\frac12^-$, $\frac32^-$ and $\frac52^-$ 
  resonance properties in the $I=0$ (empty symbols) and $I=1$ (filled
  symbols), $S=-3\ (Y=-2)$ sectors. The points with error-bars are defined
  from the masses ($M_R$) and widths ($\Gamma_R$) of the found resonances as
  $(M_R\pm\Gamma_R/2, \Gamma_R)$. The experimental $\Omega^*$ resonance masses and
  widths with their error bars are from
  refs.~\cite{PDG:2006,Biagi:1986,Aston:1987,Aston:1988}}}
\end{figure}
%
%
%
\begin{table}[ht]
\caption{Masses, widths and absolute values of coupling constants for each
  channel, in the sector $I=0$, $J^P=1/2^-$ and $S=-3$. 
 The underlining indicates open channels. 'SU(6)' stands for
  the full {\bf 35}$\times${\bf 56} result. 'SU(3)' for the {\bf
  8}$\times${\bf 8} result.}
\label{tab:1}       
\begin{tabular}{|c c|| c c c c c|} 
\hline\noalign{\smallskip}
 \multicolumn{7}{|c|}{
 $I=0,J^P=1/2^-$, 'SU(6)'  }\\
\noalign{\smallskip}\hline\noalign{\smallskip}
  ~~~M$_R$~~~ & $\Gamma_R$ & \multicolumn{5}{c|}{ $|g_i|$ }\\
 \multicolumn{2}{|c||}{ [ MeV ]} & 
\urple{$\bar{K} \Xi$} & ${\bar K}^* \Xi$ & ${\bar K}^*
\Xi^*$ & 
 $\omega\Omega$ & $\phi \Omega$  \\

\noalign{\smallskip}\hline\noalign{\smallskip}
  \CBB{1309}  &  \CBB{00} & 1.22 & 0.61 & 3.27 & 0.17 & 5.78 \\
  \CBB{1871}  &  \CBB{70} & \underline{1.57}& 3.63 & 4.17& 1.39& 3.23 \\
  \CBB{2201} &   \CBB{2.8} & \underline{0.22}& 1.63 & 2.36& 0.48 & 1.38 \\
  \CBB{2334} &   \CBB{44} & \underline{0.76} & \underline{0.37} & 0.66 & 3.44 & 1.08 \\
  \CBB{2454} &   \CBB{33} & \underline{0.38}& \underline{0.22}& \underline{0.97} & 0.18 & 4.32 \\ 
\noalign{\smallskip}\hline\noalign{\smallskip}
\urple{-}& \urple{-}&  \urple{-}& & & &  'SU(3)' \\
\noalign{\smallskip}
\noalign{\smallskip}\hline
\end{tabular}
\end{table}
%
%
\begin{table}
\caption{Same as Table \ref{tab:1} for the sector  $ I=0,~J^P=3/2^-$.}
\label{tab:2}       

\begin{tabular}{|c c||c c c c c c|} 
\hline\noalign{\smallskip}
 \multicolumn{8}{|c|}{
 $ I=0,~J^P=3/2^-$,  \CBB{'SU(6)'}}
\\
\noalign{\smallskip}\hline\noalign{\smallskip}
  M$_R$ & $\Gamma_R$ & \multicolumn{6}{c|}{ $|g_i|$ }\\
 \multicolumn{2}{|c||}{ [ MeV ]} & 
$\bar K\Xi^*$ & $\bar K^*\Xi$ & $\eta\Omega$ & $\bar K^*\Xi^*$ &
$\omega\Omega$ & $\phi\Omega$\\
\noalign{\smallskip}\hline\noalign{\smallskip}
   \CBB{1969} & \CBB{0}  & 1.78 & 2.43 & 2.37 & 1.44 & 0.00 & 2.51 \\
   \CBB{2265} & \CBB{82} & \underline{1.19} & \underline{0.34}& \underline{0.28} & 3.71 & 1.21 & 0.55 \\
   \CBB{2343} & \CBB{36} & \underline{0.42}& \underline{0.84}& \underline{0.04}& 1.01& 3.31 & 0.16 \\
   \CBB{2437} & \CBB{80} & \underline{0.07}& \underline{0.09}& \underline{1.19}& \underline{0.12}& 0.00 & 4.38 \\
\noalign{\smallskip}\hline\noalign{\smallskip}
   2051 &    \gray{ 86} & \underline{\gray{1.97}}&  & \gray{3.34}& & \multicolumn{2}{r|}{
  'SU(3)'} \\
\noalign{\smallskip}\hline
\end{tabular}
\end{table}

\begin{table}
\caption{Same as Table \ref{tab:1} for the sector  $ I=0,~J^P=5/2^-$.}
\label{tab:3}       
\begin{tabular}{|c c||c c c |} 
\hline\noalign{\smallskip}
 \multicolumn{5}{|c|}{
 $I=0,~J^P=5/2^-$, \CBB{'SU(6)'}}
\\
\noalign{\smallskip}\hline\noalign{\smallskip}
  M$_R$ & $\Gamma_R$ & \multicolumn{3}{c|}{ $|g_i|$ }\\
 \multicolumn{2}{|c||}{ [ MeV ]} & 
$\bar K^*\Xi^*$ & 
$\omega\Omega$ & $\phi\Omega$\\
\noalign{\smallskip}\hline\noalign{\smallskip}
\CBB{2376}  & \CBB{00} & 1.41 &  2.78 & 0.00
 \\
\noalign{\smallskip}\hline\noalign{\smallskip}
\end{tabular}
\end{table}
\begin{table}[h]
\caption{Same as Table \ref{tab:1} for the sector  $ I=1,~J^P=1/2^-$.}
\label{tab:4}       
\begin{tabular}{|c c|| c c c c|} 
\noalign{\smallskip}\hline\noalign{\smallskip}
 \multicolumn{6}{|c|}{
 $I=1,~J^P=1/2^-$,~\CBB{'SU(6)'} 
 }\\
\noalign{\smallskip}\hline\noalign{\smallskip}
  M$_R$ & $\Gamma_R$ & \multicolumn{4}{c|}{ $|g_i|$ }\\
 \multicolumn{2}{|c||}{ [ MeV ]} & 
\urple{$\bar{K} \Xi$} & ${\bar K}^* \Xi$ & ${\bar K}^*
\Xi^*$ & 
$\rho\Omega$  \\ 
\noalign{\smallskip}\hline\noalign{\smallskip}
   \CBB{2100} & \CBB{118} & \Red{\underline{\Black{1.47}}}& 3.39& 1.1& 2.1 \\
   \CBB{2241} & \CBB{63} & \Red{\underline{\Black{0.88}}}& \underline{0.87}& 1.67 & 3.87 \\
\noalign{\smallskip}\hline\noalign{\smallskip}
\urple{-}& \urple{-}&  \urple{-}& & \multicolumn{2}{r|}{'SU(3)'}\\
\noalign{\smallskip}\hline
\end{tabular}
\end{table}
%
\begin{table}
\caption{Same as Table \ref{tab:1} for the sector  $ I=1,~J^P=3/2^-$.}
\label{tab:5}       

\begin{tabular}{|c r||c c c c c|} 
\noalign{\smallskip}\hline\noalign{\smallskip}
 \multicolumn{7}{|c|}{
 $I=1,J^P=3/2^-$, \CBB{'SU(6)'}}
\\
\noalign{\smallskip}\hline\noalign{\smallskip}
  M$_R$ & $\Gamma_R$ & \multicolumn{5}{c|}{ $|g_i|$ }\\
 \multicolumn{2}{|c||}{ [ MeV ]} & 
$\pi\Omega$ &
$\bar K\Xi^*$ & $\bar K^*\Xi$ & $\bar K^*\Xi^*$ &
$\rho\Omega$\\
\noalign{\smallskip}\hline\noalign{\smallskip}
%
   \CBB{2018} &  \CBB{267} & \underline{2.19}& 1.64 & 2.14 & 1.75 & 0.25 \\
   \CBB{2258} &  \CBB{101} & \underline{0.54}& \underline{1.23}& \underline{0.13}& 3.19 & 2.33 \\
   \CBB{2288} &  \CBB{32} & \underline{0.33}& \underline{0.17}& \underline{0.93}& 2.29 & 3.11 \\
\noalign{\smallskip}\hline\noalign{\smallskip}
  \gray{ 2146} &    \gray{ 359} & \underline{\gray{2.30}}&  \gray{2.47}& &
  \multicolumn{2}{r|}{'SU(3)'} \\
\noalign{\smallskip}\hline
\end{tabular}
\end{table}

\begin{table}
\caption{Same as Table \ref{tab:1} for the sector  $ I=1,~J^P=5/2^-$.}
\label{tab:6}       
\begin{tabular}{|c c||c c|} 
\noalign{\smallskip}\hline\noalign{\smallskip}
 \multicolumn{4}{|c|}{
 $I=1,~J^P=5/2^-$, \CBB{'SU(6)'}}
\\
\noalign{\smallskip}\hline\noalign{\smallskip}
  M$_R$ & $\Gamma_R$ & \multicolumn{2}{c|}{ $|g_i|$ }\\
 \multicolumn{2}{|c||}{ [ MeV ]} & 
$\bar K^*\Xi^*$ & 
$\rho\Omega$ \\
\noalign{\smallskip}\hline\noalign{\smallskip}
\CBB{2324}  & \CBB{0} & 1.79 &  3.09 \\
\noalign{\smallskip}\hline
\end{tabular}
\end{table}

%
\begin{table}
\caption{Experimentally  known $\Omega^*$  
resonances ~\cite{PDG:2006,Biagi:1986,Aston:1987,Aston:1988}.
The branching ratios are relative.}
\label{tab:exp}       
\protect{
\begin{tabular}{c|c c | c |l} 
\hline\noalign{\smallskip}
 Resonance&Mass&Width&Decay&Branching \\
$I(J)^P$ & \multicolumn{2}{|c|}{ [ MeV ]} &  modes & ratios\\
\noalign{\smallskip}\hline\noalign{\smallskip}
$\Omega^*(2250)^-$&$ 2252\pm 9$&$55 \pm 18$&$\Xi^-\pi^+K^-$&
$ 1$\\
$0(?^?)$ *** & & &${\Xi^*}^0 K^-$&
$  0.7\pm 0.2$\\
\noalign{\smallskip}\hline\noalign{\smallskip}
$\Omega^*(2380)^-$&$ 2384\pm 13$&$26 \pm 23$&$\Xi^-\pi^+K^-$&
$ 1$\\
$?(?^?)$ ** & & &${\Xi^*}^0 K^-$&
$  < 0.4$\\
 & & &${\Xi}^- {\bar {K^*}}^0$&
$  0.5\pm 0.3$\\
\noalign{\smallskip}\hline\noalign{\smallskip}
$\Omega^*(2470)^-$&$ 2474\pm 12$&$72 \pm 33$&$\Omega^-\pi^+\pi^-$&
$ $\\
$?(?^?)$ ** & & & & \\
\noalign{\smallskip}\hline

\end{tabular}
}
\end{table}
\begin{acknowledgement}
We acknowledge discussions with Drs. Angels Ramos and Manuel J. Vicente-Vacas. 
This work was supported by DGI, FEDER, UE and Junta de Andaluc{\'\i}a funds 
(FIS2005-00810, HPRN-CT-2002-00311, FQM225).
\end{acknowledgement}

%

\end{document}